\documentclass[11pt]{article}
\usepackage{latexsym}
\usepackage{epsfig}

\newcommand{\beq}{\begin{eqnarray}}
\newcommand{\eeq}{\end{eqnarray}}
\newcommand{\be}{\begin{eqnarray*}}
\newcommand{\ee}{\end{eqnarray*}}

\newcommand{\ra}{\rightarrow}

\begin{document}

\centerline{\Large\bf{How I got to work with Feynman}}
\medskip
\centerline{\Large\bf{on the covariant quark model\footnote{To be published in {\it 50 Years of Quarks} by World Scientific, Singapore.}}}

\bigskip
\centerline{Finn Ravndal}
\bigskip
\centerline{\it Department of Physics, University of Oslo, Blindern, N-0316 Oslo, Norway.}

\begin{abstract}

\small{In the period 1968 - 1974 I was a graduate student and then a postdoc at Caltech and was involved with the developments of the quark and parton models. Most of this time I worked in close contact with Richard Feynman and thus was present  from the parton model was proposed until QCD was formulated. A personal account is presented how the collaboration took place and how the various stages of this development  looked like from the inside until QCD was established as a theory for strong interactions with the partons being quarks and gluons. }

\end{abstract}

Already in high school I was interested in elementary particles. Most of my knowledge stemmed from popular articles in Norwegian papers and magazines. Buzzwords like parity violation and strangeness were explained, while more abstract ideas like S-matrix or unified spinor field theory didn't make much sense except for the promise that there might be something deeper underneath. For me a particle was a track in a photographic emulsion or in the newly invented bubble chamber. 

A great inspiration was the beautiful book {\it The World of Science} where I for the first time felt that I got acquainted with modern physics and the people working in the field\cite{Watson}. In a chapter about elementary particles there was a picture of professors Feynman and Gell-Mann standing in front of a blackboard filled with diagrams and new particle symbols. They seemed very happy with what they were doing and to me it looked like an ideal occupation. This book also set Caltech on my map since it originated there.

A few years later in 1964 at the university I came across the article by Gell-Mann {\it et al.} in Scientific American about the Eightfold Way and Lie groups\cite{8fold}. It represented a real revolution - and beautiful so! The walls in my dorm room were soon filled with pictures of octets and decouplets. Here was also the prediction of the $\Omega^{-}$ which had just been discovered with the right properties. Finally there was some order in the ever increasing number of new particles which didn't seem so elementary longer. 

Around the same time came the news that this new symmetry could be explained by assuming that the particles were built up of quarks. These simple ideas explaining so much was something  I could follow. From then on I knew that the structure of elementary particles was what I wanted to study more. There were no courses in nuclear or particle physics at the technical university where I studied. But from the elegant book by Okun on weak interactions I learned a lot and it  showed me how particle physics was done in practice\cite{Okun}. The copy I had was in Russian. That was fine since I was at that time into learning a new language. That particular book meant very much to me and set me on the track I wanted to follow. 

In the summer of 1965 I was hired to work at CERN as a summer student in the Track Chamber division in a experimental group headed by B. French. My real task was to scan bubble chamber pictures for tracks coming from the decay of new particles, but soon I was asked to work more as a group theoretician than a scanner. That meant that I could spend long hours in the library reading about $SU(6)$ symmetries and the quark model. During this time I also found the book by Sakurai  very useful\cite{Sakurai}. After another year I got my engineering degree and was hired as an assistant in the physics department.

The following year I learned a lot about Lie groups and algebras from a Japanese guest professor. Knowing that the symmetry group of the 2-dimensional harmonic oscillator was $SU(2)$ and the one for the Kepler motion was $SO(4)$, I concluded from the relation $SO(4) \simeq SU(2)\times SU(2)$ that the physics of the hydrogen atom ought to be solvable in terms of the creation and annihilation operators of two 2-dimensional oscillators. Using parabolic coordinates I actually managed to show that and thus got my first publication\cite{Toyoda}. 

But it was clear that any future work on quarks and particle physics would be difficult to pursue in Norway. The summer of 1967 I spent at a school on particle physics organized at the Bohr Institute in Copenhagen. It turned out to be more of an advanced conference with current algebra as the hottest topic. To my disappointment I found most to be way over my head. But meeting other students in the same situation was a great inspiration. The following year I had to do my military service and happened to be stationed at the same research institute as Per Osland. He was in the process of applying to American universities for graduate studies. It didn't take much time to get me convinced to do the same thing. 

In the spring  of 1968 I was admitted to Caltech and left for California in August. The only scientific paper I brought with me, was a preprint from Dubna on the quark model. It presented a relativistic calculation of the magnetic moment of a proton made up of three massless quarks in a spherical cavity\cite{Dubna}. What I found so nice and fascinating, in addition to being in Russian, was the use of the Dirac equation in describing quarks. But it also showed that assigning them a new quantum number taking three different values, one could solve the spin-statistics problem. I hoped I would be able to make use of this in the coming years.

\section{Graduate student at Caltech}
During the first week at Caltech in the fall I went through two gruelling days with placement exams. As a result I also got a teachning assistantship and was assigned  to a course on intermediate quantum mechanics taught by John Bahcall. One of the first problems he gave was to estimate the flux of neutrinos from the Sun. I soon realized that physics at Caltech meant getting out measureable numbers from theoretical formulas. I got a desk in a laboratory in the Kellogg building for nuclear physics together with David Schramm. 

Initially I heard very little about quarks. But at one of the first institute colloquia the invited speaker was J. Bjorken from SLAC who talked about deep inelastic scattering and the parton model. The proton contained an infinite number of these new particles which seemed to be quarks. The atmosphere was electric. Feynman in the front row was smiling and Gell-Mann beside him was visibly excited. The old quark model seemed no longer to be relevant in this context and replaced by a more fundamental description. This was like a shock to me.

Everyone seemed to be fascinated with these new ideas but none of the graduate students or faculty I met were actively working neither on the quark nor on the parton model. In the theory group the research activity seemed to be concentrated around hadronic duality and in particular the recently proposed Veneziano scattering amplitude. 

Most of my time went to take courses. One was an introductory, theoretical course on particle physics given by Feynman himself. He started out at the simplest, phenomenological level. From data in the Particle Tables he was trying to sort out all the known particles according to quantum numbers, masses and interactions. It was obvious that he searched for some underlying order and systematics. Quarks were mentioned, but at the most rudimentary level which I already knew well. They were called A, B and C. Some of the hadronic resonances could be grouped along approximately straight Regge trajectories so that the squared masses of the particles increased linearly with their spins. This was significant, hinting at internal dynamics of constituents. Deep inelastic scattering was discussed in a lecture or two and his parton model was briefly presented. No mention of partons being quarks. The partons were just unknown, fundamental particles which he had proposed earlier in the year trying to understand the outcome of future experiments to be done at the ISR accelerator under construction at CERN.

At the end of the fall semester I realized  I needed a thesis advisor and found Steve Frautschi who tried to put me on the path of Regge theory. One of the first warm-up problems he gave me, was to generalize Yang's theorem forbidding the decay of the $\rho$-meson into two photons, to higher resonances on the $\rho$-trajectory. That was very useful because it forced me to get acquainted with the helicity formalism. Regge physics itself I did not find so tempting. But the next term I started reading more about scattering theory, Glauber description of nuclear scattering and the eikonal approximation.

During this term I followed a more advanced course on particle theory, again given by Feynman. Instead of  quantum field theory which I had expected, it was again a more phenomenological course on current commutators, infinite momentum frames and dispersion relations. Late in the spring he finished with three lectures on  deep inelastic electron scattering and the parton model. A possible connection between quarks and partons was not discussed. Instead he seemed to convey the idea that the parton model went beyond known physics and would give a completely new description of hadrons. While he had surveyed the experimental situation in particle physics  during the previous term, he now seemed to go more into theoretical aspects. The lectures were hard to follow and I could make little use of the material he presented. 

Instead I started to read up on the non-relativistic quark model on my own. In a recent paper by Faiman and Hendry they had extended the model to also describe hadron resonances by letting the quarks become orbitally excited\cite{FH}. They were assumed to be confined within a harmonic oscillator potential which had the attractive feature that explicit calculations involving the quark wave functions could easily be done analytically. From previous work I soon realized that this could be done even simpler in this symmetric quark model by using algebraic methods based on raising and lowering operators. Together with Jon Mathew who that term gave a course on advanced quantum mechanics, I reformulated the model along these lines and could soon calculate strong transition rates in a straightforward way.

During this time I became aware of the paper by Copley, Karl and Obryk from the Dalitz group in England. They used the same quark model to calculate helicity amplitudes for radiative decays rates of nucleon resonances\cite{KO}. Not only did most of the rates agree with experiments, but even the absolute signs of the different amplitudes came out right. This represented really a new, great success for the quark model. Previously it had been used primarily to explain the magnetic moments of the ground state baryons and vector mesons and the radiative transition moment for the $J=3/2$ nucleon delta resonance. The agreement with measurements was now so significant  that the quark model could no longer be just laughed away.

At the end of the semester Clem Heusch, who had given a very good course on experimental particle physics, asked me if I wanted to work at the Caltech electron synchrotron in the basement during the summer. He was heading a project within Bob Walker's experimental group where mesons were photoproduced on nuclear targets. They could measure cross-sections in different channels and were thus able to the isolate the production amplitudes for different nucleon resonances. My job was to analyze them within the quark model framework established by Copley, Karl and Obryk. Finally I was in an environment  where the quark model was at the center of most discussions!

During the summer it became clear to us that the measured photoproduction amplitudes for the nucleon resonance with mass 1720 MeV couldn't be explained if it was assigned the the known ${\bf 56}$ multiplets of the symmetry group $SU(6)$. We found instead that it could be assigned to a ${\bf 70}$ multiplet with total quark angular momentum $L = 0$.  It was the first candidate established for this more exotic multiplet. I gave a talk on these calculations at an APS meeting in Boulder late in the summer. A more general report was presented by C. Prescott at the Electron-Photon Conference at Daresbury in September. There I had my first encounter with members of the enthusiastic Dalitz group who had opened up these new applications of the quark model.  Heusch and I wrote up the results about the 1720 MeV resonance in a short paper which was published in the PRL a few months later. 

\section{Feynman and the covariant quark model}

By now I had moved in on the top floor of the new Downs-Lauritsen building with the rest of the high energy theory group. There I shared an office with a fellow graduate student, Chris Hamer from Australia. The office was just a few doors down the corridor from Feynman and Gell-Mann who bracketed the office of our beloved secretary Helen Tuck. Opposite my office resided George Zweig, the co-inventor of the quark model. Together with his foreign visitors he was  drawing beautiful quark duality diagrams which could be pushed and pulled so to give both a s- and t-channel description of a hadronic reaction from one and the same underlying diagram. That was neat and seemed to be phenomenologically useful, but this approach seemed to be far away from the quark model I felt most comfortable with. 

It was great to have followed Feynman through two advanced courses, but it didn't set me on my own line of investigations. His parton model and its applications to deep inelastic scattering which were made at SLAC, didn't seem to be much on his mind. I saw little purpose in asking him to be my thesis advisor. I had had so far no discussions with him.  More  exciting were the weekly Wednesday seminars directed by Gell-Mann. These were concentrated around scale invariance in quantum field theory in order to understand  the new experimental results coming out of SLAC.  My fellow student Rod Crewther felt at home here. But for me the seminars were often too theoretical and abstract. Feynman was usually not there and his partons were  dismissed as 'putons'. Frautschi instead suggested to think about the physics which soon would come out from the new accelerator which was getting ready at Serpukhov in Russia. Together with Chris Hamer I therefore started to consider available models for such processes which were all based on Regge theory. This approach was the best we had to describe high energy collisions and had been pursued at Caltech for many years. Several other graduate students were working on related problems under the guidance of Frautschi. Among them were Bob Carlitz and Mark Kislinger who were both one year ahead of me.

One day early in the fall Feynman was suddenly standing in the door to my office, smiling broadly as he often did. He wanted to know everything about the quark model I had used and the calculations of strong and electromagnetic decay amplitudes. It sounded like he never had heard about quarks before. I didn't know then that he had been the advisor to George Zweig when he first started to think in terms of what he then called aces as fundamental constituents of hadrons.  Someone in the synchrotron group must have told Feynman of the successes of the model. From that first encounter on we met more or less every day in the next five years. Quarks and the quark model was mostly on the agenda. Mark Kislinger was also drawn automatically into these discussions since Feynman already had established similar contacts with him.

Parallel with this very encouraging development, my work with Chris Hamer continued during the fall of 1969. Soon we realized  that we could use the eikonal method to generate higher corrections or cuts to lowest-order predictions from Regge theory. The idea was simple and elegant. It would predict cross-sections starting to increase slowly at the highest energies. Frautschi liked it and we started soon to churn out numbers for different reactions. As soon as we received the first results from Serpukhov, we saw that our model worked quite well. When I told Feynman about these calculations, he didn't show much interest. He looked more forward to the results which soon would come out from the ISR accelerator at CERN in Geneva and where he hoped to see the $1/x$ scaling in the final state as a central prediction of the parton model.

Late in the fall it was announced that Gell-Mann had received the Nobel Prize in physics for his contributions to theoretical elementary particle physics, in particular the $SU(3)$ classification scheme. No mention of George Zweig and his much more detailed quark model. It was not fair. But we had a big celebration in the new conference room. Also a Nobel Prize went to Max Delbr\"uck in biology.  A great day for Caltech and us students.

In my discussions with Feynman we had become convinced that the quark model had to include a coupling to mesons which we called the  'Mitra term' after the Indian physicist who had previously studied it\cite{MR}. It was necessary in order to describe strong decays where the momentum of the final-state meson is small. This new term depended on the internal quark momentum and would therefore not be suppressed in these cases. From my earlier quark investigations I knew that it would automatically appear if a meson coupled to another hadron through the divergence of the four-dimensional axial vector current. Feynman found this so significant that he suddenly showed up one day with the first ideas for what later would be called the covariant quark model. While Mark had previously shown little interest in the old quark model, he was now suddenly all excited. We now had a Lorentz-invariant Hamiltonian giving the squared masses of hadrons in terms of the Dirac four-momenta of the quarks and the squares of their invariant separation. It was thus a relativistic generalization of the non-relativistic, harmonic oscillator quark model. By minimal coupling to external vector and axial vector fields, we could then isolate the hadronic vector and axial vector currents expressed in terms of quark variables. It was compact in its description and we found it to be very elegant. One nice feature Feynman stressed, was that it didn't assume any explicit values for the quark masses. For a ground state baryon they came automatically out to be one third of the baryon mass. We could early see that it would automatically reproduce many of the features of the non-relativistic model we needed to have in place. But all its consequences had now to be worked out in detail and checked against measurements.

Early in 1970 Feynman had gone to a meeting on the East Coast and reported on my work with Chris on the eikonalized Regge model for the Serphukov total cross sections without telling me in advance. On the first day back at Caltech he came to my office, very excited. He told me that on the plane back to LA he had been thinking about these collisions in terms of his parton picture and had come up with a very simple model which predicted that the total cross-sections had to increase as the squared logarithm of the collision energy when this went into the asymptotic region.  This simple behaviour seemed to be consistent with the observations. All hadronic total cross-sections should approach the same asymptotic value. He sketched the derivation on the blackboard. In the center-of-mass frame for the collision, it is the partons with the smallest momenta and therefore with a known distribution, which will dominate the process. So one can express the total cross-section in terms of parton-parton cross-sections. But each parton is again made up of other partons so that the energy dependence of a parton-parton scattering must be the same as for the hadronic process we started with. This bootstrap idea is similar to what later could be formulated in terms of the renormalization group. As a result he could derive an integral equation for the cross-section which can then be shown to have the above-mentioned analytical solution. The meaning was that we should study this model more carefully later. But we never came back to it. We were too busy with the covariant quark model. Some twenty years later I presented this derivation at a meeting at Yale\cite{Yale}. 

During the spring of 1970 together with Hamer and Frautschi we finished two papers on the eikonalization of Regge amplitudes and sent them off to the Physical Review\cite{Eikonal}. Late in the spring I was invited by G. Chew up to Berkeley to give an institute colloquium about these calculations.  All the local big shots were in the audience and I felt greatly honored. But it was not the physics which my heart was beating for. I received \$50 to cover my expenses and took the Pacific Coast Highway back to Los Angeles and Pasadena.  

Together with Kislinger and Feynman I had during the same time used our covariant quark model to calculate every known strong and electromagnetic amplitude which had been measured or could be of interest. We had discovered that the model had a serious flaw related to the removal of un-physical degrees of freedom. It resulted in a form factor which was just wrong and showing up in all transitions.  We had to replace it with a more {\it ad hoc} or phenomenological choice which we just called the 'fudge factor'. There were so many other dramatic assumptions in the model, that we didn't worry too much about this particular point. What we liked and gave us confidence in this rough construction, was the following properties:

\begin{enumerate}
\item It was covariant so that calculations could be done in any Lorentz frame.
\item  No explicit quark masses appeared in the model.
\item  Excited hadrons were on straight Regge trajectories.
\item  The vector current was conserved in the symmetric limit and the axial current satisfied PCAC.
\end{enumerate}

\begin{figure}
  \begin{center}
    \epsfig{figure=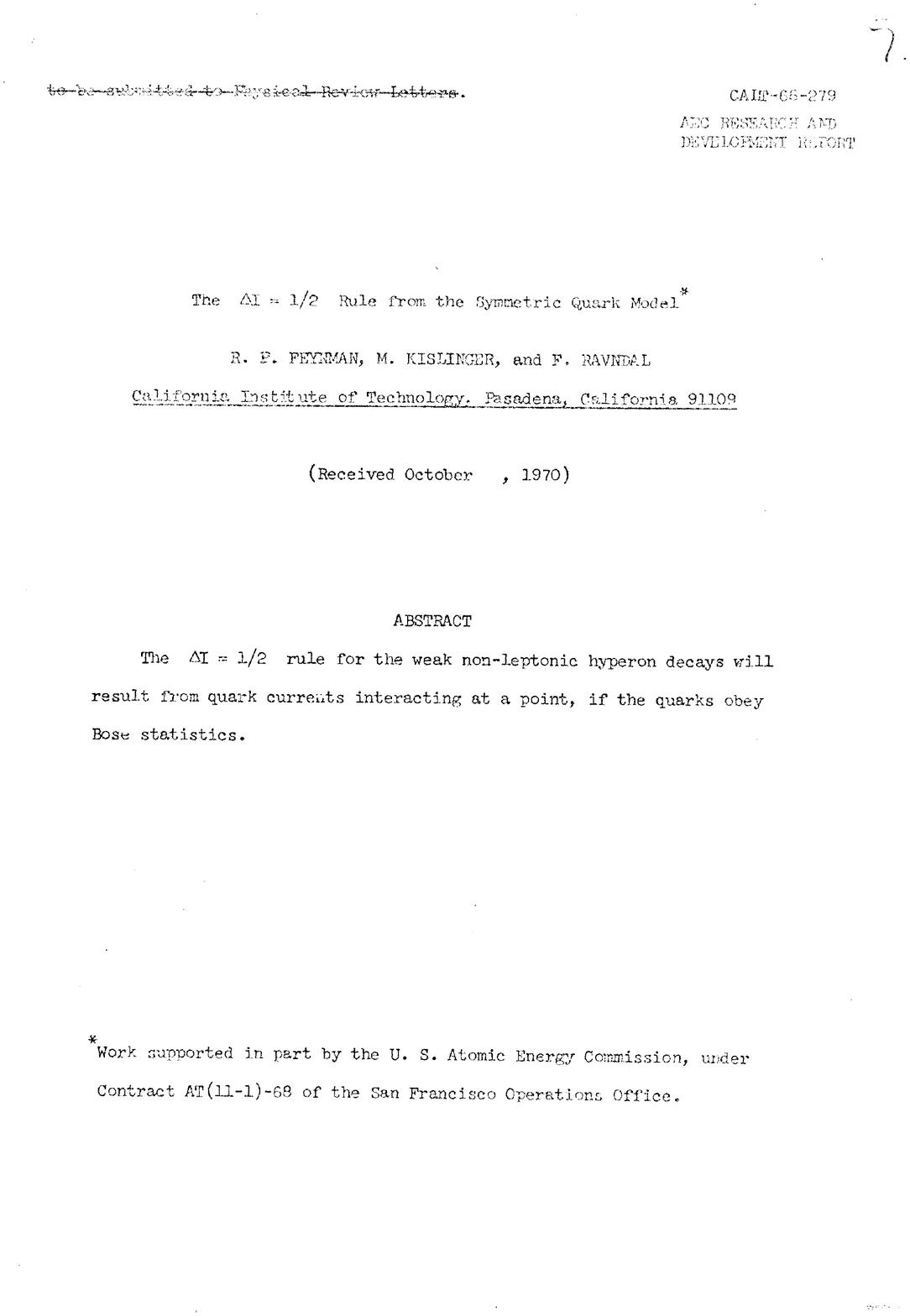, height = 200mm}
  \end{center}           
  \caption{\footnotesize Front page of preprint where quarks are suggested to be bosons.}
  \label{Fig.1}
\end{figure}
No other quark model had these desirable properties. Needless to say, it also reproduced all the good results of the non-relativistic model. We had nothing new to say about the spin-statistics problem. Feynman liked to say that the Pauli theorem relating spin and statistics only applied to free particles and therefore not to permanently confined quarks. Thus he joked that they could just as well be bosons and there would be no problem. Gell-Mann in his seminars had already discussed both parastatistics and the color quantum number of Han and Nambu. We felt this was a much deeper problem which would be solved at a later time. 
But one day we realized that if quark really had symmetric statistics as for bosons, the $I = 1/2$ selection rule in non-leptonic, weak decays would be explained. Feynman became very excited, wrote together a short letter and submitted it to PRL\cite{FKR-bosons}. But a couple of weeks later we decided to withdraw it since we had been informed that the same observation had been made previously by others. The arguments were later included in our main article about the covariant quark model.

In the fall it became clear that we had enough results and Feynman wanted to write it all up. We were all three happy with the results. He suggested to write the first version himself. I checked the formulas and added in the references. In order to check the numerical calculations,  Feynman had gotten hold of a small Wang office computer in the Downs building. He had learned to program it in some cryptic language and loved to see all the numbers come out.  The final paper was sent off to the Physical Review just before Christmas 1970\cite{FKR-1971}. 

Although very few else around us took much interest in what we had done, this two-year long, concentrated effort by Feynman had been necessary for him to become convinced about the physical existence of quarks. We realized certainly that our version of the  quark model was just a better crutch for that purpose and would not have a lasting impact. But for Feynman the quarks themselves were from then on real and he was now open to identify them with his proposed partons. 

At the start of the following term I was told by Frautschi that I would be able to graduate in the spring with a thesis based on the quark model. In order to finish off with something new, I extended the covariant quark model to calculate radiative transitions with off-shell photons. The resulting amplitudes could be measured in the electroproduction of nucleon resonances\cite{electro}.

On February 9, 1971 Southern California was rocked by the strong San Fernando earthquake. That afternoon Feynman was scheduled to give his first, regular seminar about the parton model at Caltech. The bulding was shaking from aftershocks which didn't seem to worry him while the rest of us were pretty nervous.  This was the first time that his partons were officially identified with quarks. From now on it would be the quark-parton model as already explored at SLAC for more than two years. The same day Harald Fritzsch had also arrived in Pasadena to work with Gell-Mann.

Some weeks later Feynman asked me if I wanted to stay on for another two years in a postdoc position starting in the fall. I was elated, hoping that I could then devote myself to the parton model instead. It was obvious that it  represented the future. In addition, this new position meant that I didn't have to go back to Norway or find a new job some other place.

\section{Postdoc at Caltech}

The next  Electron-Photon Conference took place at Cornell in August, 1971. Feynman went there and wanted me to come with him. During the meeting he was highly sought out by people who wanted to see him and discuss with him. Sometimes I felt that the main reason for me being there was to act like a 'bodyguard' for him. From the talks we became more and more convinced that quarks and partons were behind everything that was presented. Feynman was very happy.  During a talk on what was then called the dual parton model which we never had understood, Feynman pushed his elbow into my side when he saw on a slide that the model could even explain a small bump around 3 GeV in the $\mu^+\mu^-$  invariant mass in new Drell-Yan cross-section data. He didn't believe it. Two years later this bump became the $J/\psi$ particle.

On the plane back to Los Angeles, we encountered bad weather and Feynman became visibly nervous. This was in stark contrast to his careless appearance during the earthquake half a year earlier. As a distraction from this bumpy ride he wanted that we should plan our activities at Caltech in the fall. They would be concentrated around a new lecture course he would give on the evidence for quarks and partons. For every lecture he would prepare notes which I then should correct, elaborate and make publishable in a joint book. In addition he wished to terminate a project with a graduate student who was working on a QED problem.

It didn't take many weeks before I concluded that I couldn't continue on the book project. The first reason was that I felt that I already was much too close to Feynman and identified with him in almost  everything I did. To a large degree that was the actual situation and I had to establish some kind of independence, even if he had asked me to work with him. Secondly, when I read his notes with his own words and explanations, I also felt  it like a sacrilege to the rest of the physics community if I should start to modify his own formulations. He accepted that and the job was given to another graduate student, Arturo Cisneros. He also let the notes essentially intact and the book was published the following year\cite{PhotonHadron}. I myself skipped to a large extent the lectures since I felt that I knew it all. Instead I finished a paper on electroproduction of nucleon resonances in the quark model\cite{electro} and started to consider new applications to diffractive and weak production in neutrino scattering\cite{neutrino}.

At this time the light-cone current algebra of Fritzsch and Gell-Mann seemed to give a better theoretical understanding of the scaling seen in deep inelastic electron scattering\cite{lightcone}. Feynman himself didn't think so and showed little interest. The main reason was the fundamental mass scale which was implicit in his model and manifested itself for instance in the transverse momentum cut-off of the hadrons in the final state. The current commutators on the light cone were abstracted from free quark theory and did not contain such a quantity. From the over-all structure of this description applied to inclusive scattering it was clear that it could reproduce many of the parton model results. This was in particular demonstrated by my colleagues Tony Hey and Jeff Mandula\cite{HeyMandula}. 

In my own discussions with Feynman we tried instead to extend our quark model so to give a more symmetric description of hadronic couplings. Previously this had been done via PCAC where we could replace a meson field with the divergence of the axial current. We had in mind the dual diagrams which George Zweig and collaborators had considered. John Schwarz had by that time joined the theory group. In his string theory such couplings could easily be constructed. Together with the straight Regge trajectories in that description we thought that there could be some connection to our harmonic quark model, but we didn't pursue this much further.

When Feynman continued his lectures in the spring of 1972 he extended the parton model to describe the final hadronic state in terms of quark fragmentation functions. This opened up a new chapter of applications and we started to talk about jets in the final state. Together with two postdocs Mike Gronau and Yair Zarmi I started discussions about how these ideas could be tested in deep inelastic scattering experiments. Later in the year this was written up in joint paper\cite{GronauZarmi}.

During that term Feynman went to Hungary to take part in a neutrino conference at Balatonf\"ured. He came back fired up with the first quantitative experimental confirmation of the parton model and the fractional quark charges. This was the measurement of the famous factor of $5/18$ which related the deep inelastic electron scattering cross section to the corresponding cross section with neutrinos. 
Around this time he had started to think that the charge of a single quark could be measured directly by summing up the charges from the jet fragments in the final state. This possibility caught the attention of Glennys Farrar and Jon Rosner\cite{FarrarRosner}. In the fall I travelled to Fermilab with Feynman where we among others met Stan Brodsky who also had considered this proposal in his collaboration with Farrar. He argued convincingly that such a measurement would not be possible. Feynman realized that he had been wrong and was visibly shaken afterwards.

In the meantime Murray Gell-Mann and Harald Fritzsch had returned from CERN where they had spent the last year. The quarks in the light-cone algebra had now three colours. In addition to solving the spin-statistics problem in hadronic spectroscopy, Rod Crewther had shown from the Adler anomaly of the axial current that the decay $\pi^0 \ra \gamma\gamma$ got the right rate when multiplied by this factor of three. It also meant that the cross section for electron-positron annihilation into hadrons at high energies would be three times larger than what followed from just adding up the three squared quark charges which were expected to contribute. But both the light-cone approach and the parton model said that the ratio of this cross section to that of annihilation into a muon pair, should be constant with increasing energy. Instead the available data showed that the ratio continued to rise with increasing energy. These results seemed to undermine our whole understanding.

Feynman didn't seem to be so worried. His parton model was an attempt to give a framework for high energy hadronic reactions built upon as few assumptions as possible. There was no Lagrangian or particular field theory behind, but just the most basic properties of special relativity and quantum mechanics. And with this extra, built-in mass scale the underlying theory had to represent really new and different physics not contained in conventional quantum field theory. This became especially clear in a discussion between Feynman and Fritzsch about the final state in hadronic electron-positron annihilation. While Feynman talked about and drew pictures of two hadronic jets coming out in the directions of the produced quarks,  Fritzsch argued instead for having hadrons in all directions.

At the Batavia conference in September 1972 Gell-Mann presented his new understanding of strong interactions worked out together with Fritzsch and J. Bardeen during the year at CERN. The three new colour charges represented an invariance under local gauge transformations governed by the group $SU(3)$. Observed hadrons should be colour singlets described by a Yang-Mills field theory with eight coloured gluons. Such a field theory had many times earlier come up as a possibility in Gell-Mann's seminars, most often in connection with chiral invariance. But this time it was much more convincing. It could be the definite theory of strong interactions with a structure very similar to QED. It was already christened QHD for Quantum Hadrodynamics but was renamed QCD the year after.

The question was now what one could do with it. It was a beautiful proposal, but in many ways too good to be true. How could it be compatible with the parton model and in particular with Feynman's description of quark fragmentation and jets which the now standard light-cone formulation said very little about? Feynman himself found it interesting but remained sceptical. For instance, in a field theory with gluons it would be difficult or impossible to recover the simple separation of quark distribution functions in the initial state and quark fragmentation functions in the final state which was so important for phenomenological applications of the parton model. Similar problems would arise in Drell-Yan production and the final state in $e^+e^-$ annihilation. We stuck or clung to the belief that the parton model was something else than standard quantum field theory.

I continued with the quark model and finished also a couple of papers on the parton model. In one I showed that there should be no azimuthal dependence in the hadronic distribution of deep inelastic scattering resulting from quark fragmentation in a jet\cite{azimuth}, a result which survived the later and more accurate treatment in QCD. But even if no one around me started to work on Yang-Mills theories, it was certainly discussed. And the use of them in electroweak unification got more attention for the first time. Feynman had never shown any particular interest in these developments. When he once was asked why, he answered that one could not discover new physics in the electroweak sector from the requirement of making a theory renormalizable. It sounded like he thought that also in this sector one would need a more radical approach. After these comments there would be nothing more said about unified theories. On my own I tried to orient myself and wanted to consider the possibility of seeing charmed quarks in current-induced reactions at high energies. Together with a graduate student I wrote a paper about this within the parton model\cite{Novoseller}. When I told Feynman about our results, he got almost upset. He saw no reason to introduce new degrees of freedom in a situation where there still were so much uncertainty.

My two years as a postdoc were now running out. In the meantime I had already secured a permanent job at my old university in Norway. But Feynman meant that I could find a better job at a more exciting place. And to my great surprise he could a short time afterwards tell me that he had secured money for me to stay another year in my postdoc position. That was a relief, but it also meant that I didn't have a permanent job to return to the year after.

Late in the spring 1973 we heard that the beta-function for QCD had been calculated on the East Coast and found to be negative, implying asymptotic freedom. The implications for deep inelastic scattering we already knew from the lecture notes of Sidney Coleman. My first reaction was great exhilaration since suddenly the treatment of partons as free particles could be understood. But Feynman showed little or no interest in this result. That was surprising since his parton model now had a field-theoretic formulation. One reason was the unsettling situation with the total $e^+e^-$ cross section for which the latest experiments at the Cambridge accelerator still gave values much larger than expected. The factor of three due to the new colours didn't seem to be the solution.

Back at Caltech in the fall we in the younger generation realized that renormalization and the calculation of Feynman diagrams would be necessary in order to participate in the exploration of  the new QCD. But this was a direction of particle physics for which we were not prepared, in spite of having Feynman and Gell-Mann around us on a daily basis. There had been very little or no quantum field theory in standard courses with the weight instead on more phenomenological aspects. In one of his Wednesday seminar Gell-Mann wanted to discuss the renormalization group and its use in QCD. It was not of much help and we felt disappointed, expecting more from one of the originators of this fundamental method.

Instead of going into all the new and detailed calculations having to do with applications of QCD, I stuck with the original parton model and all its predictions which seemed to be confirmed in an increasing number of new experiments. I got engaged in a new project together with Thom Curtright and Jeff Mandula which we called the covariant operator parton model\cite{COPM}. This new formulation enabled us to derive Feynmans results more systematically based on the properties of the quarks fields entering the current matrix elements for the different processes. It gave us a certain satisfaction although we knew that it was most likely doomed. In particular, there were no gluon degrees of freedom and therefore also no gluon jets.

Feynman didn't take much interest in what we were doing, but Gell-Mann asked me to give a seminar about it. What instead caught Feynman's attention was 2-dimensional QED and its use as an illustration of the parton model\cite{CKS}. It gave exact scaling and could also describe the fragmentation in the final state. In addition, the partons were permanently confined by a linear potential. We had earlier been reminded about the attractiveness of this potential for the quark model by Ken Kauffmann who was graduate student at the time when we worked with harmonic confinement. He had pointed out that if the potential was due to some underlying field theory, then it ought to be linear since only in that case does it have a simple Fourier transform, namely $1/Q^4$. Some years later this idea would be a central part of the dual quark model as formulated by Fred Zachariassen and his collaborators\cite{Zach}.

In December that year we drove down in Feynmans car to the particle physics meeting at Irvine. The first experimental results were presented confirming the small scaling violations in deep inelastic electron scattering which should follow from QCD. The original parton model seemed to some of us to be defeated and we worried even more for the implied lack of factorization between initial and final states which were so central to its successes. But Feynman was unperturbed, partly because the  $e^+e^-$ cross section was still not understood. At the meeting Abdus Salam even proposed that it showed the existence of an electromagnetic Pomeron, an idea Gell-Mann in the audience found laughable. We headed back to Pasadena and Xmas vacation, confused and not knowing that this great bewilderment would be gone in less than a year.

My last term at Caltech in the spring 1974 went primarily into finishing up older projects and preparing for the future. I got a position by Nordita in Copenhagen and looked forward to going back to Europe after six wonderful years at Caltech. In this period I had been so fortunate to be present when QCD was established as a possible fundamental theory for strong interactions. But in order to verify it and extract its experimental consequences one would need methods of quantum field theory. Many of us felt we were ill equipped in this area with the emphasis on phenomenology that we had had. New results and insights were soon coming out from groups on the East Coast where they were better prepared in pure theory. A certain gloom settled in the last months I was there. Some of us felt that our group had lost the leading r\^ole we previously had felt we had. Half jokingly it was said that the only contribution from Caltech that would remain, would be the $u,d$ and $s$ quark names and not the script $p,n$ and $\lambda$ used until then by these new QCD theoreticians. 

On one of my last days in Pasadena I was sitting in the conference room where so much of the new physics had been presented and discussed in the previous years. With me was Steve Frautschi who also had felt that the group had lost some of its momentum because of this new direction in particle physics opening up. And that was perhaps a paradox. Feynman and Gell-Mann were to a large degree the founders of modern particle physics, but now it seemed that a new breed of younger people with a different background would be needed there. It was time for me to leave Caltech.

\section{Feynman seen a little from the inside}

I never understood why I should be the one of the many students and postdocs at Caltech that Feynman chose to engage himself  with in such a close way for five years in this period. It started obviously out from his desire to get a fast start on the evidence for quarks and the quest to understand their relation to his own partons. But starting out as an ordinary student-professor collaboration, it changed soon also into a personal friendship which was to a large extent independent of a scientific collaboration. In some ways it could have been similar to an old-fashioned assistantship where the assistant fulfils a personal need of having someone around and always available for more practical chores. I would think that several earlier and later collaborators of Feynman have had similar experiences.

One of the tasks which I soon realized was to some extent my responsibility, was to keep him informed about what was happening, what kind of ideas were discussed and what was new in preprints and other publications. In his office there were usually very few papers and journals to see and essentially no books. That surprised me and certainly gave the impression that he was independent of the works of others. And to a large extent that was probably also true. He loved to work out every calculation himself, to understand every problem in his own way. This inclination was also reflected in the way he prepared his lectures. Usually he was then sitting in his office an hour or two before each lecture and writing out in detail every calculation and argument he needed even if he had given the same course the year before or earlier. Everything had to be fresh in his mind and worked out from scratch. In spite of this unusual effort, he never seemed to get stressed when it had do with teaching.

That he didn't read the books and works of others, was to some degree just a show. Early on in our collaboration I was asked to 'babysit' his two kids Carl and Michelle. I was certainly curious about how the great man lived and used the occasion to sneak around in his house. From his large living room there was a stair down to a cozy room in the basement, filled with books and scientific literature around the walls. That was a relief to see and made him a little less godlike. He later told me that this room was his 'Cave'.

At the same time he probably worked out much on his own what already was known in the literature. One morning he came directly to my office and was all excited about a new derivation he had discovered to explain the so-called 'twin paradox' in the theory of relativity. It involved two observers with identical radio emitters exchanging signals as one travels away from Earth and later returns. Using the standard, relativistic Doppler-effect he could then easily illustrate how time evolves differently for the two observers. What Feynman had found was nothing other than the K-calculus of H. Bondi\cite{Bondi}. Later I was inclined to think that he had arrived at this delightful insight through possible consulting work at JPL and one of their planetary satellite programs.

He was a very kind person, and almost often very happy and content. I never heard an angry word from him or a disparaging remark. He never made you feel inferior in any way although he was way above me and most others. Instead he could care about others and helped out when he could. Probably the first time I saw this side of him, was when he got me X-rayed. We went almost daily to lunch together, usually to the 'Greasy' student cafeteria. One day we passed one of these buses that travel around and offer examinations of peoples lungs. With a short account of how he had lost his first wife to tuberculosis and that he didn't want to loose me the same way, he almost pushed me into the bus and the X-ray machine. I couldn't protest but my lungs had probably preferred to avoid all this radiation. 

Sometimes when he went away from Pasadena, he wanted me to go with him. On one of our first visits to Fermilab he was asked on a short notice by our experimental colleague Barry Barish to come and check out a potential discovery of a wrong-sign lepton in one of the first Caltech-run neutrino experiments there. We arrived late in the afternoon and drove directly to the primitive quonset hut where the data was analyzed. It was dark, rainy and muddy. In all respects very different from the clean facilities available today. After a short look at the measurements Feynman concluded that the signal was just a fluke and nothing to care about. At that time there was much talk of possible heavy leptons which in some theories should show up in these new experiments.

At a later visit to Fermilab we were invited to dinner by the director Bob Wilson. In the afternoon we drove over to his house that lay on a small hill out in the countryside. It was in the winter and the narrow road up there was icy and slippery. The car spun and Wilson came out and helped me push it to the house with Feynman at the wheel. This set the tone for the rest of the evening which took place in the kitchen with the three of us. Much of the talk was about all times and colleagues. When it touched upon the lives of Bob Serber and Kitty Oppenheimer, I felt like I was back in war-time Los Alamos.

It was also at Fermilab I for the first time saw a weaker and more vulnerable side of Feynman. This was in the evening after he had been told that he had made a mistake in his argument about measuring the quark charges from summing up the hadronic charges in a jet. He had afterwards become unusual quiet and withdrawn. After we had gone to bed in the guest apartment we disposed, he called me into his room and wanted to tell me about what was important in life and how one should avoid wasting time on meaningless endeavours. It was like a father-to-son talk for which I was not prepared and didn't feel had some much to do with me. 

It was probably this other side I also got a glimpse of on the flight back to Los Angeles after the Cornell meeting in 1971. I had not expected to see him nervous in the plane because of bad weather outside. He, the most rational man on Earth! But sometimes he a could show such contradictory sides of himself. 
\begin{figure}[h]
  \begin{center}
    \epsfig{figure=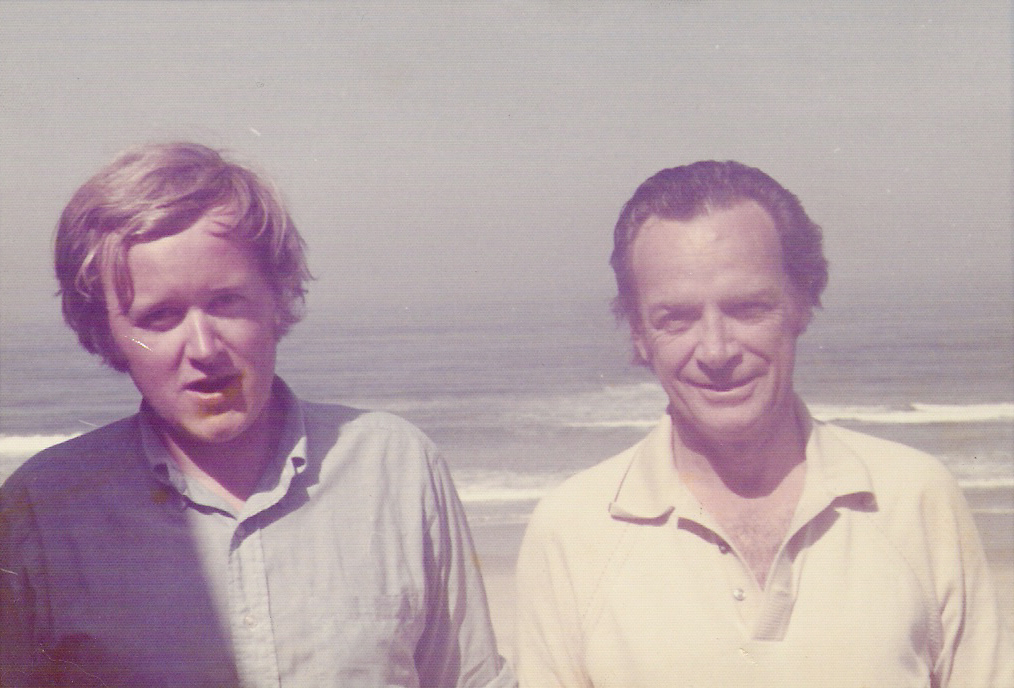, height = 80mm}
  \end{center}           
  \caption{\footnotesize On the beach at Feynman's summer house in Mexico.}
  \label{Fig.1}
\end{figure}

One weekend he had invited me and my wife to his summerhouse at the beach south of Tijuana in Mexico. We went in his car and he was at the wheel. Down there the roads were pretty narrow and curvy which he didn't seem to notice. Instead he drove rather fast and totally careless, as we saw it. This time I was the one who was scared while he was  just happily grinning and talking. But as soon as we had safely arrived, everything around this scary experience was soon forgotten and we had a wonderful weekend with the family, including his sister and mother.

Many years later I came across an interview with Freeman Dyson\cite{Dyson}. There he told about a similar experience with Feynman already in 1947 when they drove from Cornell to a seminar in Rochester. That car ride also frightened Dyson a lot and shows that this streak in Feynman's character was already there when we experienced it on the road in Mexico. What it showed, I will probably never understand. 

Through Feynman I was also so happy to meet many other famous physicists. Once I was with my wife invited to his house in Altadena in connection with a visit to Caltech by his colleague and PhD advisor John Wheeler. It was clear that Feynman had deep respect for Wheeler who in turn at that setting was dominated by an unusually assertive wife which reduced my own impression of the great physicist. But Wheeler made it up again the following day in a beautiful lecture on general relativity.

Most likely it was also Feynman who in 1971 set me in contact with Max Delbr\"uck who wanted a house sitter for the summer months he was in Cold Spring Harbor or in Europe. Delbr\"uck was very interested in what we could explain with the quark model. But the result was that he wanted me later to transfer to a postdoc position in his own group to work on the fungus {\it Phycomyces}. I'm happy I didn't accept the offer. Even so we were allowed to stay in his beautiful house the next three summers. It was also by the Delbr\"ucks that W. Heisenberg stayed when he came to visit Caltech in 1974. While Gell-Mann kept away that day, Feynman was the considerate host. But at dinner in the evening Feynman became very direct in his criticism of the talk Heisenberg had given earlier in the day\cite{Heisenberg}. This side of Feynman I never saw myself, but I knew how important it was for him to sort out bad from good physics.

In the spring of 1974 he told me that he would give the commencement speech at the graduation ceremony at Caltech some weeks later. He already  had some ideas around honesty in science that he wanted to emphasize and showed me a preliminary manuscript to read. Some of the arguments I had heard before, but didn't realize how much attention this talk about {\it Cargo Cult Science} would later generate. I didn't even attend the ceremony and was instead preparing my departure from Caltech after six wonderful years there. Looking back, there is much I would have done differently had I again been given the opportunity to be close to this great physicist and human being.

\section{Epilogue}

I came back on short visits to Calfornia several times in the following years. Already in 1975 I met him at a meeting at Stanford. We then had lunch together with L. Alvarez who was very happy for just having debunked a possible discovery of a magnetic monopole. Feynman was  diagnosed with cancer around this time. During a talk at a meeting at Caltech on QCD in 1978 it seemed to me that he had lost some of the strength and confidence he earlier usually showed. On the last day the final talk was given by Gell-Mann who pointed out that Feynman just had turned 60 years. He ended by thanking him explicitly for his many contributions which had made the success of this new physics possible. I was sitting next to Feynman and know that this open acknowledgement from his colleague was much appreciated. 

When I met Feynman the following year, he was just out of the hospital and I had to visit him at home in his own bedroom. As ten years earlier I was surprised to see that also here he had books all around the room. Next to his bed lay a copy of Newton's {\it Principia}. That we didn't discuss. Instead he was very excited about his new, personal computer. It was a PET which stood on a nearby desk. He was very proud of a couple of programs he ran for me and which recursively generated some very nice patterns. He always had loved computing and computers.

During all these years I stayed in contact with Helen Tuck who was Feynman's secretary and kept me informed about what happened. When I was on a visit to Santa Barbara in 1984, she called me and said that Feynman wanted to see me. Again in some way I couldn't accept or believe it, and declined with the excuse of not having enough time. Four years later I was at the University of Illinois in Urbana-Champaign and heard Feynman's voice out in the corridor happily talking physics. He was so alive!  But what I heard was just some tapes the students played of his old lectures. They had just been informed that he had died the day before. 

A few years ago I browsed through the book Feynman's  daughter Michelle wrote about his many letters\cite{Michelle}. One is to a friend on the East Coast who is asked if he could come up with some money for a Norwegian postdoc who planned to go back for a job which he didn't find  so attractive. Apparently Feynman got the money and I thereby my sixth year there. My postdoc arrangement had been against department traditions and regulations and came out just because of Feynman's expressed wish. After two years it should be over, but again he reached out a helping hand. I never understood what really happened.

\end{document}